\newcommand\apjcls{1}
\newcommand\aastexcls{2}
\newcommand\othercls{3}

\newcommand\papercls{\aastexcls}
\documentclass[tighten, times, twocolumn, trackchanges]{aastex61}

\if\papercls \apjcls
\usepackage{apjfonts}
\else\if\papercls \othercls
\usepackage{epsfig}
\fi\fi
\usepackage{ifthen}
\usepackage{natbib}
\usepackage{amssymb, amsmath}
\usepackage{appendix}
\usepackage{etoolbox}
\usepackage[T1]{fontenc}
\usepackage{paralist}

\if\papercls \apjcls
\newcommand\aas{\ref@jnl{AAS Meeting Abstracts}}
\newcommand\dps{\ref@jnl{AAS/DPS Meeting Abstracts}}
\newcommand\maps{\ref@jnl{MAPS}}
\else\if\papercls \othercls
\usepackage{astjnlabbrev-jh}
\fi\fi

\if\papercls \aastexcls
\hypersetup{citecolor=blue, 
            linkcolor=blue, 
            menucolor=blue, 
            urlcolor=blue}  
\else
\usepackage[
bookmarks=true,           
bookmarksnumbered=true,   
colorlinks=true,          
citecolor=blue,           
linkcolor=blue,           
menucolor=blue,           
urlcolor=blue,            
linkbordercolor={0 0 1},  
pdfborder={0 0 1},
frenchlinks=true]{hyperref}
\fi

\if\papercls \othercls
\newcommand{\eprint}[1]{\href{http://arxiv.org/abs/#1}{#1}}
\else
\renewcommand{\eprint}[1]{\href{http://arxiv.org/abs/#1}{#1}}
\fi

\providecommand{\adsurl}[1]{\href{#1}{ADS}}

\makeatletter
\patchcmd{\NAT@citex}
  {\@citea\NAT@hyper@{%
     \NAT@nmfmt{\NAT@nm}%
     \hyper@natlinkbreak{\NAT@aysep\NAT@spacechar}{\@citeb\@extra@b@citeb}%
     \NAT@date}}
  {\@citea\NAT@nmfmt{\NAT@nm}%
   \NAT@aysep\NAT@spacechar\NAT@hyper@{\NAT@date}}{}{}

\patchcmd{\NAT@citex}
  {\@citea\NAT@hyper@{%
     \NAT@nmfmt{\NAT@nm}%
     \hyper@natlinkbreak{\NAT@spacechar\NAT@@open\if*#1*\else#1\NAT@spacechar\fi}%
       {\@citeb\@extra@b@citeb}%
     \NAT@date}}
  {\@citea\NAT@nmfmt{\NAT@nm}%
   \NAT@spacechar\NAT@@open\if*#1*\else#1\NAT@spacechar\fi\NAT@hyper@{\NAT@date}}
  {}{}
\makeatother

\makeatletter
\DeclareRobustCommand{\lowcase}[1]{\@lowcase#1\@nil}
\def\@lowcase#1\@nil{\if\relax#1\relax\else\MakeLowercase{#1}\fi}
\makeatother

\DeclareSymbolFont{UPM}{U}{eur}{m}{n}
\DeclareMathSymbol{\umu}{0}{UPM}{"16}
\let\oldumu=\umu
\renewcommand\umu{\ifmmode\oldumu\else\math{\oldumu}\fi}
\newcommand\micro{\umu}
\if\papercls \othercls
\newcommand\micron{\micro m}
\else
\renewcommand\micron{\micro m}
\fi
\newcommand\microns{\micron}

\let\oldsim=\sim
\renewcommand\sim{\ifmmode\oldsim\else\math{\oldsim}\fi}
\let\oldpm=\pm
\renewcommand\pm{\ifmmode\oldpm\else\math{\oldpm}\fi}
\newcommand\by{\ifmmode\times\else\math{\times}\fi}
\newcommand\ttt[1]{10\sp{#1}}

\newbox{\wdbox}
\renewcommand\c{\setbox\wdbox=\hbox{,}\hspace{\wd\wdbox}}
\renewcommand\i{\setbox\wdbox=\hbox{i}\hspace{\wd\wdbox}}

\newcount\timect
\newcount\hourct
\newcount\minct
\newcommand\now{\timect=\time \divide\timect by 60
         \hourct=\timect \multiply\hourct by 60
         \minct=\time \advance\minct by -\hourct
         \number\timect:\ifnum \minct < 10 0\fi\number\minct}

\catcode`@=11

\newcommand\comment[1]{}

\newcommand\commenton{\catcode`\%=14}

\renewcommand\math[1]{$#1$}
\newcommand\mathshifton{\catcode`\$=3}

\let\atab=&
\newcommand\atabon{\catcode`\&=4}

\let\oldmsp=\sp
\let\oldmsb=\sb
\def\sp#1{\ifmmode
           \oldmsp{#1}%
         \else\strut\raise.85ex\hbox{\scriptsize #1}\fi}
\def\sb#1{\ifmmode
           \oldmsb{#1}%
         \else\strut\raise-.54ex\hbox{\scriptsize #1}\fi}
\newbox\@sp
\newbox\@sb
\def\sbp#1#2{\ifmmode%
           \oldmsb{#1}\oldmsp{#2}%
         \else
           \setbox\@sb=\hbox{\sb{#1}}%
           \setbox\@sp=\hbox{\sp{#2}}%
           \rlap{\copy\@sb}\copy\@sp
           \ifdim \wd\@sb >\wd\@sp
             \hskip -\wd\@sp \hskip \wd\@sb
           \fi
        \fi}
\def\msp#1{\ifmmode
           \oldmsp{#1}
         \else \math{\oldmsp{#1}}\fi}
\def\msb#1{\ifmmode
           \oldmsb{#1}
         \else \math{\oldmsb{#1}}\fi}

\def\supon{\catcode`\^=7}

\def\subon{\catcode`\_=8}

\def\supsubon{\supon \subon}

\newcommand\actcharon{\catcode`\~=13}

\newcommand\paramon{\catcode`\#=6}

\comment{And now to turn us totally on and off...}

\newcommand\reservedcharson{ \commenton  \mathshifton  \atabon  \supsubon 
                             \actcharon  \paramon}

\catcode`@=12
\reservedcharson

\if\papercls \apjcls

\else

\fi

\newcommand\Webb{{\em James Webb Space Telescope}}
\newcommand\JWST{{\em JWST}}
\newcommand\chisq{\ifmmode{\chi\sp{2}}\else\math{\chi\sp{2}}\fi}
\newcommand\redchisq{\ifmmode{ \chi\sp{2}\sb{\rm red}}
                    \else\math{\chi\sp{2}\sb{\rm red}}\fi}
\newcommand\Teq{\ifmmode{T\sb{\rm eq}}\else$T$\sb{eq}\fi}
\newcommand\mjup{\ifmmode{M\sb{\rm Jup}}\else$M$\sb{Jup}\fi}
\newcommand\rjup{\ifmmode{R\sb{\rm Jup}}\else$R$\sb{Jup}\fi}
\newcommand\msun{\ifmmode{M\sb{\odot}}\else$M\sb{\odot}$\fi}
\newcommand\rsun{\ifmmode{R\sb{\odot}}\else$R\sb{\odot}$\fi}
\newcommand\mearth{\ifmmode{M\sb{\oplus}}\else$M\sb{\oplus}$\fi}
\newcommand\rearth{\ifmmode{R\sb{\oplus}}\else$R\sb{\oplus}$\fi}

\shorttitle{Line-transition Data Compression}
\shortauthors{Patricio Cubillos}

\begin{document}

\title{An Algorithm to Compress Line-transition Data for
  Radiative-transfer Calculations}

\author{Patricio~E.~Cubillos}
\affiliation{Space Research Institute, Austrian Academy of Sciences,
              Schmiedlstrasse 6, A-8042, Graz, Austria}

\email{patricio.cubillos@oeaw.ac.at}

\begin{abstract}
  Molecular line-transition lists are an essential ingredient for
  radiative-transfer calculations.  With recent databases now
  surpassing the billion-lines mark, handling them has become
  computationally prohibitive, due to both the required processing
  power and memory.  Here I present a temperature-dependent algorithm
  to separate strong from weak line transitions, reformatting the
  large majority of the weaker lines into a cross-section data file,
  and retaining the detailed line-by-line information of the fewer
  strong lines.  For any given molecule over the 0.3--30 {\micron}
  range, this algorithm reduces the number of lines to a few million,
  enabling faster radiative-transfer computations without a
  significant loss of information.  The final compression rate depends
  on how densely populated is the spectrum.  I validate this algorithm
  by comparing Exomol's HCN extinction-coefficient spectra
  between the complete (65 million line transitions) and compressed
  (7.7 million) line lists.  Over the 0.6--33 {\micron} range, the
  average difference between extinction-coefficient values is less
  than 1\%.  A Python/C implementation of this algorithm is
  open-source and available
  at \href{https://github.com/pcubillos/repack}
  {https://github.com/pcubillos/repack}.  So far, this code handles
  the Exomol and HITRAN line-transition format.
\end{abstract}


\section{INTRODUCTION}
\label{introduction}

The study of exoplanet atmospheres and their spectra critically
depends on the available laboratory and theoretical data of gaseous
species
\citep{FortneyEtal2016whiteLabData}.  The discovery of highly
irradiated sub-stellar atmospheres has motivated the
compilation of molecular line-transition lists at temperatures
far above those of the Earth atmosphere
\citep[][]{RothmanEtal2010jqsrtHITEMP, TennysonEtal2016jmsExomol}.
However, these newest databases are starting to grow into the
$\sim$billions of line transitions
\citep[e.g.,][]{RothmanEtal2010jqsrtHITEMP,
  YurchenkoEtal2011mnrasNH3opacities,
  YurchenkoTennyson2014mnrasExomolCH4}.

To date, medium- to low-resolution multi-wavelength observations of
exoplanets cover a broad wavelength range ($\sim$0.3 to 30 {\micron}),
requiring the use of the line-transition data nearly in their
entirety.  With the arrival of future facilities, like the {\Webb}
({\JWST}), this picture will remain.  Such large line-transition data
files render radiative-transfer calculations computationally
prohibitive, both in terms of the necessary memory and processing
power.

To keep the molecular line lists manageable, authors commonly set a
fixed opacity cutoff, discarding all lines weaker than a certain
threshold \citep[e.g.,][]{SharpBurrows2007apjOpacities}.  However,
this approach is at best below optimal, as one could remove entire
absorption bands at certain wavelengths, or retain a large
number of line-transitions that do not significantly contribute to the
opacity.

Inspired by the idea of \citet{HargreavesEtal2015apjHotCH4} of
separating line-by-line and continuum line-transition information, I
devised an algorithm to reduce the amount of line-transition data
required for radiative-transfer calculations, with minimal loss of
information.  Using this approach, one retains the full information
only of the stronger lines that dominate the absorption spectrum, and
compresses the combined information of the many-more weak lines into a
cross-section data file, as a function of wavenumber and temperature.
Since, the interpretation of mid- and low-resolution observations
relies more on the total opacity contribution rather than the
individual line transitions, this algorithm allows for a significant
performance improvement of radiative-transfer calculations.

\section{METHODS}
\label{sec:methods}

The integrated absorption intensity (or opacity or extinction
coefficient) of a line transition (in cm$\sp{-1}$) can be expressed as
\begin{equation}
\label{eq:intensity}
\small
S\sb{j} = \frac{\pi e\sp{2}}{m\sb{e}c\sp{2}} \frac{(gf)\sb{j}}{Z\sb{i}(T)}
          n\sb{i} \exp\left(-\frac{h c E\sb{\rm low}\sp{j}}{k\sb{B}T}\right)
            \left\{1-\exp\left(-\frac{hc\nu\sb{j}}{k\sb{B}T}\right)\right\},
\end{equation}
where $gf\sb{j}$, $\nu\sb{j}$, and $E\sp{j}\sb{\rm low}$ are the
weighted oscillator strength, central wavenumber, and lower-state
energy level of the line transition $j$, respectively; $Z\sb{i}$ and
$n\sb{i}$ are the partition function and number density of the isotope
$i$, respectively; $T$ is the temperature; $e$ and $m\sb{e}$ are the
electron's charge and mass, respectively; $c$ is the speed of light,
$h$ is Planck's constant; and $k\sb{\rm B}$ is the Boltzmann's
constant.

For any given molecule, the number density of an isotope can be
written as $n\sb{i} = n\sb{m} q\sb{i}$, where $n\sb{m}$ is the
molecule number density and $q\sb{i}$ is the isotopic abundance
fraction (which can be assumed to be at Earth values).
Then, by knowing the set of isotopic fractions for a given
molecule, we can express the line intensities per unit of the
molecule's number density (in cm$\sp{2}$molec$\sp{-1}$) as $s\sb{j} =
S\sb{j}/n\sb{m}$.

To compute the extinction-coefficient spectrum, one needs to broaden
each line according to the Voigt profile function (the convolution of a
Doppler and a Lorentz profile), and then summing the contribution from
all lines.  Thus, to identify the dominant lines, one has to consider
the dilution of the line intensity by the Voigt broadening.

\begin{figure}[t]
\centering
\includegraphics[trim=18 8 40 34, width=\linewidth, clip]{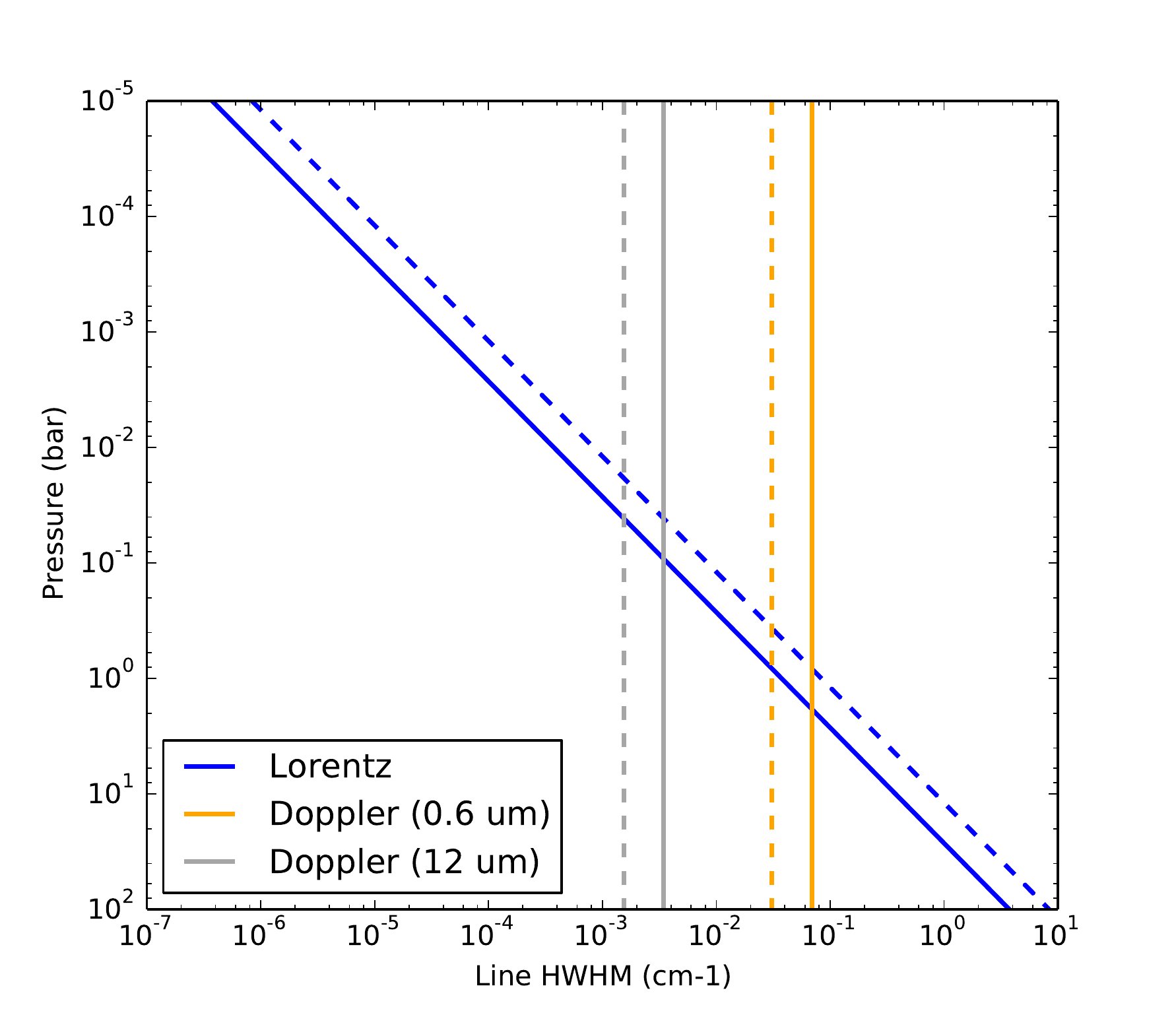}
\caption{
Doppler and Lorentz profile half-width at half-maximum (HWHM) for HCN
in an typical gaseous exoplanet atmosphere.  The solid and dashed
lines denote the HWHM at 2500 and 500 K, respectively.  Typically,
transmission spectroscopy probes pressures between 1 and
$\ttt{-3}$ bar, whereas day-side emission probes between 10 and
$\ttt{-2}$ bar.  The Lorentz HWHM proportionality with pressure sets
the major variation over an atmosphere, increasing exponentially with
depth.}
\label{fig:broadening}
\end{figure}

In practice, for a given molecule, the Voigt broadening profile varies
weakly over neighboring lines.  Since the Doppler broadening is
simpler to compute than the Lorentz profile, I approximate the
Voigt broadening by the Doppler broadening profile:
\begin{equation}
\label{eq:doppler}
I\sb{{\rm D},j}(\nu) = \frac{1}{\delta\sb{{\rm D},j}\sqrt{\pi}}
  \exp\left(-\frac{(\nu - \nu\sb{j})\sp{2}}{\delta\sb{{\rm D},j}\sp{2}}\right), 
\end{equation}
with line width
\begin{equation}
\delta\sb{{\rm D},j} = \frac{\nu\sb{j}}{c} \sqrt\frac{2 k\sb{B} T}{m\sb{i}},
\end{equation}
where $m\sb{i}$ is the mass of the isotope.

Figure \ref{fig:broadening} shows the typical values for the
Doppler and Lorentz half-width at half-maximum (HWHM) for exoplanet
atmospheres.  The Doppler broadening dominates the line profile above
{\sim}0.3 bar, across temperature ranges of 500 to 2500~K and
wavelengths of 0.6 to 12~{\microns}.  In this pressure range, the
line-transition widths range between {\sim}$\ttt{-3}$ cm$\sp{-1}$ and
{\sim}$\ttt{-1}$ cm$\sp{-1}$, depending on the wavelength and
atmospheric parameters.
\citet[][submitted]{CubillosEtal2017aaRadiusPressure} studied the
atmospheric pressures probed by transmission and emission spectroscopy
of gaseous exoplanets, for a wide range of planetary properties.
They found that the typical optical-to-infrared transmission
observations probe pressures between 1 and $\ttt{-3}$ bar, whereas
day-side emission observations probe between 10 and $\ttt{-2}$ bar.
Therefore, most of the observable atmosphere is under the Doppler
broadening regime.
If one approximates the lineshape by the Doppler profile,
Eq.~(\ref{eq:doppler}) tells then that the maximum intensity of a line
is approximately $s'\sb{j} = s\sb{j}/\delta\sb{\rm D}\sqrt{\pi}$
(hereafter, called diluted line intensity).  In
Section \ref{sec:validation} we show that this is an acceptable
assumption, even when there is significant Lorentz broadening, for
medium or low-resolution observations.

\subsection{Line-flagging Algorithm}

To efficiently identify dominant from weak lines, one can start by
selecting the strongest line in a given wavelength range, say line
transition $j$, and compute its Doppler profile $s\sb{j} I\sb{{\rm
D},j}(\nu\sb{k})$.
Then, one flags out the surrounding lines whose diluted intensity
($s'\sb{k}$) is smaller than the profile of line $j$, with a threshold
tolerance $f\sb{\rm tol}$, i.e.:
\begin{equation}
s'\sb{k} < f\sb{\rm tol}\,s\sb{j}I\sb{{\rm D},j}(\nu\sb{k}).
\end{equation}

\begin{figure}[t]
\centering
\includegraphics[trim=15 0 30 25, width=\linewidth,clip]{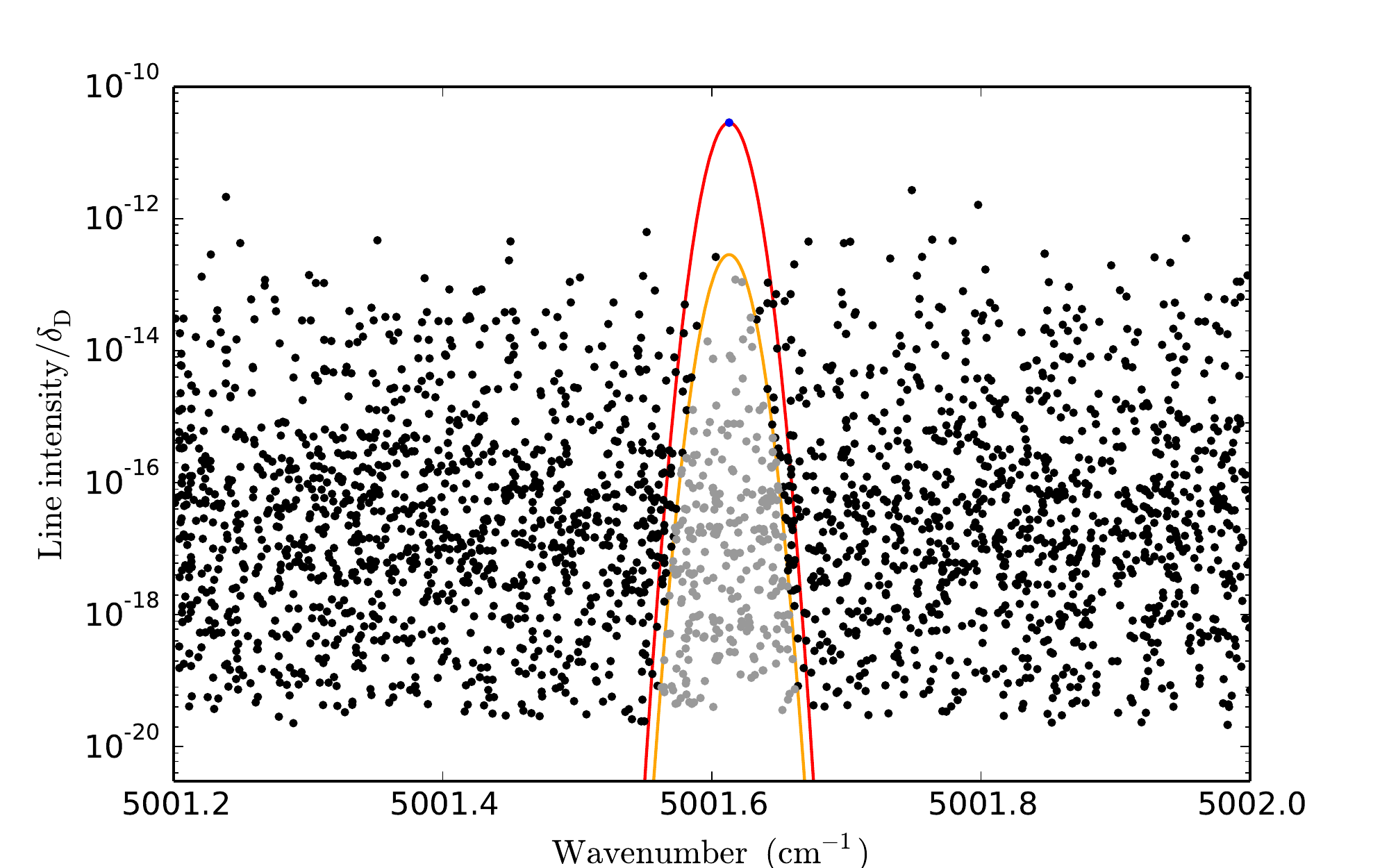}
\caption{
Detail of the line flagging.  Each dot represents the diluted
intensity of HCN line transitions over a narrow wavelength range.  The
red curve shows the Doppler broadening profile of the strongest line
in this range.  The gray dots denote the weak line transitions, whose
diluted intensity is smaller than the Doppler curve with a 
threshold tolerance of $f\sb{\rm tol}=0.01$ (orange curve).}
\label{fig:flagging}
\end{figure}

Since the line intensity decays exponentially as one moves away from
its center, the flagged (weak) lines do not significantly contribute
to the extinction coefficient.  This effectively avoids the need to
broaden most of the lines in a database.  Line intensities span several
orders of magnitude, and thus, only the few strongest lines at any
given wavelength dominate the absorption spectrum (Figure
\ref{fig:flagging}).  The algorithm then proceeds with the next
un-flagged strongest line transition, and so on.

Now, different lines can dominate the spectrum depending on the
atmospheric temperature (see Eq.\ (\ref{eq:intensity})).  To account
for this, one repeats the flagging process at the two extreme
temperatures to consider (a user choice, for example, 300 and 3000 K
is appropriate for sub-stellar objects).

After one identifies strong and weak lines, one preserves the full
line-by-line information only for the strong lines (for radiative
transfer, the $\nu\sb{j}$, $E\sb{\rm low}\sp{j}$, $gf\sb{j}$, and the
isotope ID suffice).  The information from the large majority of weak
lines can be compressed into a continuum extinction coefficient table
as function of wavelength and temperature. To avoid broadening each
line, one simply add the line intensity to the nearest tabulated
wavenumber point, diluting the line according to the wavenumber
sampling rate \citep[following][]{SharpBurrows2007apjOpacities}.

\section{Open-source Implementation}

Along with this article, I provide an open-source version of this
algorithm (under the MIT license), available
at \href{https://github.com/pcubillos/repack}
{https://github.com/pcubillos/repack}.  This is a Python package
(accelerated with C subroutines) compatible with Python2 and Python3,
running on both Linux and OSX.  This package handles the Exomol and
HITRAN input line-transition formats.  The routine's
performance, varies with the size of the initial database, the number
of evaluated profiles (which depends on $f\sb{\rm tol}$), and how
densely packed are the line transitions.  For an Intel Core i7-4790
3.60 GHz CPU, the routine runs the Exomol HCN
\citep[65 million lines,][]{HarrisEtal2006mnrasHCN, HarrisEtal2008mnrasHCN13,
BarberEtal2014mnrasExomolHCN},
NH$\sb{3}$ 
\citep[$\sim$1 billion,][]{YurchenkoEtal2011mnrasNH3opacities},
and CH$\sb{4}$ \citep[10
billion,][]{YurchenkoTennyson2014mnrasExomolCH4} databases in {\sim}10
minutes, {\sim}7 hours, and {\sim}5 days, respectively.  The
performance scales somewhat faster than linear with the number
of line transitions.  Certainly, the gain of working with the
compressed line lists more than compensates for the time spent running
this routine (a one-time run) for the largest data bases.
Ultimately the line-by-line compression rate will depend on how
compact or saturated is a line list.  For example, with a threshold
tolerance of 0.01, this algorithm compresses the Exomol HCN database
by $\sim$90\% in the 0.6--33 {\micron} range.  The algorithm
compresses the denser Exomol NH$\sb{3}$ line list ($\sim$1 billion) by
$\sim$95\%.

\begin{figure*}[t]
\centering
\includegraphics[trim=12 6 15 5, width=0.92\linewidth, clip]{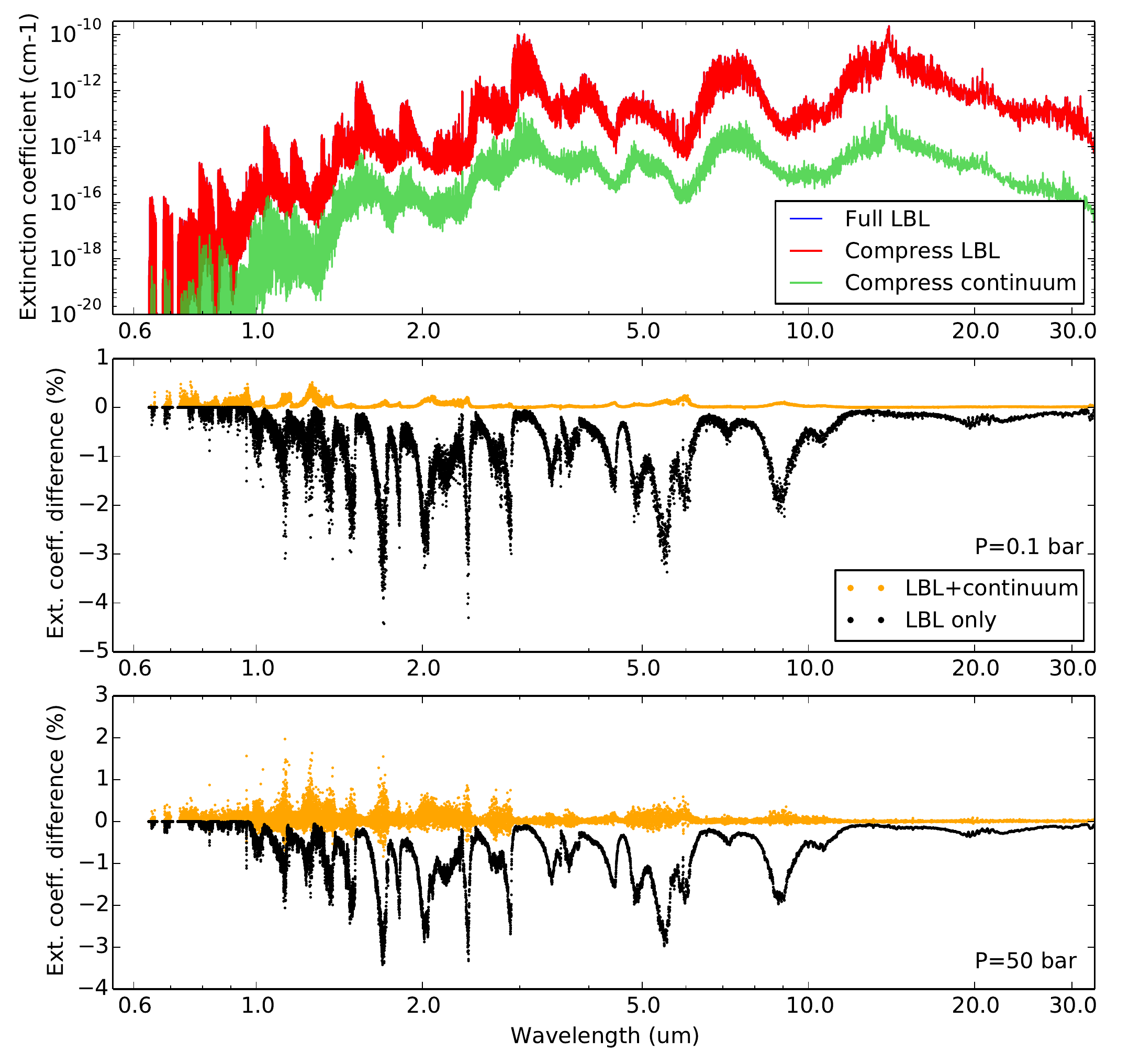}
\caption{{\bf Top:} HCN extinction coefficient from the full Exomol
  line-list (blue), the compressed line-by-line (LBL) dataset (red),
  and the compressed continuum data (green) at 0.1 bar and 1540
  K.  Note that, at this scale the full and compressed LBL curves are
  indistinguishable, and thus, the red curve falls on top of the blue
  curve.  The spectra in this panel are sampled over a regular
  wavenumber grid with 0.1 cm$\sp{-1}$ sampling rate, without
  instrumental broadening.  {\bf Middle:} HCN extinction-coefficient
  spectrum difference between the full and compressed line lists at
  0.1 bar and 1540 K.  The orange and black curves show the
  difference using the total (LBL plus continuum) components and only
  the LBL component of the compressed line list, respectively.  These
  spectra simulate an instrumental resolving power of 1~cm$\sp{-1}$
  (Gaussian filter applied after the radiative-transfer calculation).
  {\bf Bottom:} same as the middle panel, evaluated at 50 bar.}
\label{fig:HCNspectra}
\end{figure*}

\subsection{Validation}
\label{sec:validation}

To validate the compression algorithm, I compare the
extinction-coefficient spectra produced from the compressed and the
complete line-by-line database for HCN from
Exomol.  This data base comprises $\sim$65
million lines over the 0.6--33.0 {\micron} range.  Adopting a
threshold tolerance of $f\sb{\rm tol}=0.01$, and temperature values
between 500 and 3000 K, the algorithm retains 7.7 million lines.

To compute the HCN's extinction-coefficient spectrum, I use the Python
Radiative-transfer in a Bayesian framework package \citep[Pyrat
Bay\footnote{\href{http://pcubillos.github.io/pyratbay}
{http://pcubillos.github.io/pyratbay}},][in
prep.]{CubillosEtal2017apjPyratBay}.  Pyrat Bay is based on the
Bayesian Atmospheric Radiative Transfer
package \citep{Blecic2016phdThesis, Cubillos2016phdThesis}.
Figure \ref{fig:HCNspectra} (top panel) shows the resulting spectra
for a typical Jupiter-composition planet, at an atmospheric
temperature of 1540 K and pressure of 0.1 bar.
Figure~\ref{fig:HCNspectra} (middle panel) shows the difference in
extinction coefficient between the full line-list and the compressed
dataset.  The spectra for this panel simulate an instrumental
resolving power of 1~cm$\sp{-1}$ (approximately the highest resolution
that the {\it JWST} instruments will achieve).  The LBL component
alone reproduces the full line-list extinction coefficient down to
$\sim$1\% on average, and down to a few percent in the worst case.
When, considering both the LBL and continuum components, the
compressed dataset reproduces the full line-list spectrum well under
1\%.

At higher pressures, where the Lorentz profile dominates the line
broadening, the compressed dataset still reproduces well the full
line-list spectrum (Fig.~\ref{fig:HCNspectra}, bottom panel).  Some
over-estimated values arise because the compressed continuum opacity
does not consider the Lorentz broadening.  However, the differences
are still on the order of a few percent.
The mismatch varies with the given instrumental resolution,
with a coarser resolution producing smaller differences.\\

In summary, I presented an efficient compression algorithm that
identifies the line transitions that dominate a spectrum.  This
algorithm is aimed to serve radiative-transfer modeling of medium to
low spectral resolution data over broad wavelength ranges, like that
of the {\JWST}.

\acknowledgments

I thank Dr. L. Fossati and Dr. J. Blecic for useful
comments.  I also thank the anonymous referee for his/her time
and valuable comments.  I thank contributors to
Numpy \citep{vanderWaltEtal2011numpy}, SciPy
\citep{JonesEtal2001scipy}, Matplotlib
\citep{Hunter2007ieeeMatplotlib}, the Python Programming Language,
the developers of the AASTeX latex
template \citep{AASteam2016aastex61}; and the free and open-source
community.  I acknowledge the Austrian
Forschungsf{\"o}rderungsgesellschaft FFG projects ``RASEN'' P847963
and ``TAPAS4CHEOPS'' P853993, the Austrian Science Fund (FWF) NFN
projects S11607-N16 and S11604-N16, and the FWF project
P27256-N27. The Reproducible Research Compendium (RRC) of this
article is available at
\href{https://github.com/pcubillos/Cubillos2017\_repack}
{https://github.com/pcubillos/Cubillos2017\_repack}.

\software{
{Repack: \href{https://github.com/pcubillos/repack}
              {https://github.com/pcubillos/repack}},
{the Python Radiative Transfer in a Bayesian framework:
 \href{http://pcubillos.github.io/pyratbay}
      {http://pcubillos.github.io/pyratbay}},
and Latex template: \href{https://github.com/pcubillos/ApJtemplate}
                      {https://github.com/pcubillos/ApJtemplate}.
}


\begin{thebibliography}{18}
\expandafter\ifx\csname natexlab\endcsname\relax\def\natexlab#1{#1}\fi

\bibitem[{{AAS Journals Team} \& {Hendrickson}(2016)}]{AASteam2016aastex61}
{AAS Journals Team}, \& {Hendrickson}, A. 2016, AASJournals/AASTeX60: Version
  6.1

\bibitem[{{Barber} {et~al.}(2014){Barber}, {Strange}, {Hill}, {Polyansky},
  {Mellau}, {Yurchenko}, \& {Tennyson}}]{BarberEtal2014mnrasExomolHCN}
{Barber}, R.~J., {Strange}, J.~K., {Hill}, C., {Polyansky}, O.~L., {Mellau},
  G.~C., {Yurchenko}, S.~N., \& {Tennyson}, J. 2014, \mnras, 437, 1828,
  \adsurl{http://adsabs.harvard.edu/abs/2014MNRAS.437.1828B},
  \eprint{1311.1328}

\bibitem[{{Blecic}(2016)}]{Blecic2016phdThesis}
{Blecic}, J. 2016, ArXiv e-prints,
  \adsurl{http://adsabs.harvard.edu/abs/2016arXiv160402692B},
  \eprint{1604.02692}

\bibitem[{{Cubillos} {et~al.}(2017{\natexlab{a}}){Cubillos}, {Blecic}, \&
  {Harrington}}]{CubillosEtal2017apjPyratBay}
{Cubillos}, P., {Blecic}, J., \& {Harrington}, J. 2017{\natexlab{a}}, in prep.

\bibitem[{{Cubillos} {et~al.}(2017{\natexlab{b}}){Cubillos}, {Kubyshkina},
  {Fossati}, {Mordasini}, \& {Lendl}}]{CubillosEtal2017aaRadiusPressure}
{Cubillos}, P., {Kubyshkina}, D., {Fossati}, L., {Mordasini}, C., \& {Lendl},
  M. 2017{\natexlab{b}}, submitted.

\bibitem[{{Cubillos}(2016)}]{Cubillos2016phdThesis}
{Cubillos}, P.~E. 2016, ArXiv e-prints,
  \adsurl{http://adsabs.harvard.edu/abs/2016arXiv160401320C},
  \eprint{1604.01320}

\bibitem[{{Fortney} {et~al.}(2016){Fortney}, {Robinson}, {Domagal-Goldman},
  {Sk{\aa}lid Amundsen}, {Brogi}, {Claire}, {Crisp}, {Hebrard}, {Imanaka}, {de
  Kok}, {Marley}, {Teal}, {Barman}, {Bernath}, {Burrows}, {Charbonneau},
  {Freedman}, {Gelino}, {Helling}, {Heng}, {Jensen}, {Kane}, {Kempton},
  {Kopparapu}, {Lewis}, {Lopez-Morales}, {Lyons}, {Lyra}, {Meadows}, {Moses},
  {Pierrehumbert}, {Venot}, {Wang}, \& {Wright}}]{FortneyEtal2016whiteLabData}
{Fortney}, J.~J. {et~al.} 2016, ArXiv e-prints,
  \adsurl{http://adsabs.harvard.edu/abs/2016arXiv160206305F},
  \eprint{1602.06305}

\bibitem[{{Hargreaves} {et~al.}(2015){Hargreaves}, {Bernath}, {Bailey}, \&
  {Dulick}}]{HargreavesEtal2015apjHotCH4}
{Hargreaves}, R.~J., {Bernath}, P.~F., {Bailey}, J., \& {Dulick}, M. 2015,
  \apj, 813, 12, \adsurl{http://adsabs.harvard.edu/abs/2015ApJ...813...12H},
  \eprint{1510.06982}

\bibitem[{{Harris} {et~al.}(2008){Harris}, {Larner}, {Tennyson}, {Kaminsky},
  {Pavlenko}, \& {Jones}}]{HarrisEtal2008mnrasHCN13}
{Harris}, G.~J., {Larner}, F.~C., {Tennyson}, J., {Kaminsky}, B.~M.,
  {Pavlenko}, Y.~V., \& {Jones}, H.~R.~A. 2008, \mnras, 390, 143,
  \adsurl{http://adsabs.harvard.edu/abs/2008MNRAS.390..143H},
  \eprint{0807.0717}

\bibitem[{{Harris} {et~al.}(2006){Harris}, {Tennyson}, {Kaminsky}, {Pavlenko},
  \& {Jones}}]{HarrisEtal2006mnrasHCN}
{Harris}, G.~J., {Tennyson}, J., {Kaminsky}, B.~M., {Pavlenko}, Y.~V., \&
  {Jones}, H.~R.~A. 2006, \mnras, 367, 400,
  \adsurl{http://adsabs.harvard.edu/abs/2006MNRAS.367..400H},
  \eprint{astro-ph/0512363}

\bibitem[{Hunter(2007)}]{Hunter2007ieeeMatplotlib}
Hunter, J.~D. 2007, Computing In Science \& Engineering, 9, 90

\bibitem[{Jones {et~al.}(2001)Jones, Oliphant, Peterson,
  {et~al.}}]{JonesEtal2001scipy}
Jones, E., Oliphant, T., Peterson, P., {et~al.} 2001, {SciPy}: Open source
  scientific tools for {Python}, [Online; accessed 2017-02-12]

\bibitem[{{Rothman} {et~al.}(2010){Rothman}, {Gordon}, {Barber}, {Dothe},
  {Gamache}, {Goldman}, {Perevalov}, {Tashkun}, \&
  {Tennyson}}]{RothmanEtal2010jqsrtHITEMP}
{Rothman}, L.~S. {et~al.} 2010, \jqsrt, 111, 2139,
  \adsurl{http://adsabs.harvard.edu/abs/2010JQSRT.111.2139R}

\bibitem[{{Sharp} \& {Burrows}(2007)}]{SharpBurrows2007apjOpacities}
{Sharp}, C.~M., \& {Burrows}, A. 2007, \apjs, 168, 140,
  \adsurl{http://adsabs.harvard.edu/abs/2007ApJS..168..140S},
  \eprint{arXiv:astro-ph/0607211}

\bibitem[{{Tennyson} {et~al.}(2016){Tennyson}, {Yurchenko}, {Al-Refaie},
  {Barton}, {Chubb}, {Coles}, {Diamantopoulou}, {Gorman}, {Hill}, {Lam},
  {Lodi}, {McKemmish}, {Na}, {Owens}, {Polyansky}, {Rivlin}, {Sousa-Silva},
  {Underwood}, {Yachmenev}, \& {Zak}}]{TennysonEtal2016jmsExomol}
{Tennyson}, J. {et~al.} 2016, Journal of Molecular Spectroscopy, 327, 73,
  \adsurl{http://adsabs.harvard.edu/abs/2016JMoSp.327...73T},
  \eprint{1603.05890}

\bibitem[{van~der Walt {et~al.}(2011)van~der Walt, Colbert, \&
  Varoquaux}]{vanderWaltEtal2011numpy}
van~der Walt, S., Colbert, S.~C., \& Varoquaux, G. 2011, Computing in Science
  \& Engineering, 13, 22

\bibitem[{{Yurchenko} {et~al.}(2011){Yurchenko}, {Barber}, \&
  {Tennyson}}]{YurchenkoEtal2011mnrasNH3opacities}
{Yurchenko}, S.~N., {Barber}, R.~J., \& {Tennyson}, J. 2011, \mnras, 413, 1828,
  \adsurl{http://adsabs.harvard.edu/abs/2011MNRAS.413.1828Y},
  \eprint{1011.1569}

\bibitem[{{Yurchenko} \&
  {Tennyson}(2014)}]{YurchenkoTennyson2014mnrasExomolCH4}
{Yurchenko}, S.~N., \& {Tennyson}, J. 2014, \mnras, 440, 1649,
  \adsurl{http://adsabs.harvard.edu/abs/2014MNRAS.440.1649Y},
  \eprint{1401.4852}
\end{thebibliography}

\end{document}